\documentclass[prl,aps,twocolumn,showpacs,superscriptaddress,amsmath,amssymb]{revtex4}

\usepackage{graphicx}
\usepackage{psfrag}
\usepackage{epsfig}
\usepackage{color}
\usepackage{subfigure}
\definecolor{dred}{rgb}{0.7,0.0,0.0}
\usepackage{dcolumn}
\usepackage{bm}

\begin{document}
\title{Transport anisotropy of the pnictides studied via \\
Monte Carlo simulations of a Spin-Fermion model}

\author{Shuhua Liang}
\affiliation{Department of Physics and Astronomy, University of Tennessee, Knoxville, Tennessee 37996, USA}
\affiliation{Materials Science and Technology Division, Oak Ridge National Laboratory, Oak Ridge, Tennessee 37831, USA}
\author{Gonzalo Alvarez}
\affiliation{Computer Science and Mathematics Division and Center for Nanophase Materials Sciences,
Oak Ridge National Laboratory, Oak Ridge, Tennessee 37831, USA}
\author{Cengiz \c{S}en}
\affiliation{Department of Physics and Astronomy, University of Tennessee, Knoxville, Tennessee 37996, USA}
\affiliation{Materials Science and Technology Division, Oak Ridge National Laboratory, Oak Ridge, Tennessee 37831, USA}
\author{Adriana Moreo}
\affiliation{Department of Physics and Astronomy, University of Tennessee, Knoxville, Tennessee 37996, USA}
\affiliation{Materials Science and Technology Division, Oak Ridge National Laboratory, Oak Ridge, Tennessee 37831, USA}
\author{Elbio Dagotto}
\affiliation{Department of Physics and Astronomy, University of Tennessee, Knoxville, Tennessee 37996, USA}
\affiliation{Materials Science and Technology Division, Oak Ridge National Laboratory, Oak Ridge, Tennessee 37831, USA}

\date{\today}

\begin{abstract}
An undoped three-orbital Spin-Fermion model for the Fe-based superconductors
is studied via Monte Carlo techniques in two-dimensional clusters. 
At low temperatures, the magnetic and one-particle spectral 
properties are in agreement with neutron and photoemission 
experiments. Our main results are the
resistance vs. temperature curves that display the same features
observed in BaFe$_2$As$_2$ detwinned single 
crystals (under uniaxial stress), 
including a low-temperature anisotropy
between the two directions followed by
a peak at the magnetic ordering temperature, that qualitatively appears related
to short-range spin order and concomitant Fermi Surface orbital order. 
\end{abstract}

\maketitle


{\it Introduction.} 
In early studies of Fe-based superconductors~\cite{review},
it was widely assumed that Fermi Surface (FS) nesting 
was sufficient to understand the undoped-compounds magnetic order
with wavevector ${\bf Q}$ = ($\pi$,0)~\cite{cruz} and the
pairing tendencies upon doping. Neutron scattering reports 
of spin-incommensurate order~\cite{spinIC} 
are in fact compatible with the nesting scenario. 
However, several recent experimental results cannot be explained by FS nesting, including
{\it (i)} electronic ``nematic'' tendencies
in Ca(Fe$_{1-x}$Co$_x$)$_2$As$_2$~\cite{davis}; 
{\it (ii)} orbital-independent
superconducting gaps~\cite{shimojima} found using laser angle-resolved photoemission (ARPES)
spectroscopy on BaFe$_2$(As$_{0.65}$P$_{0.35}$)$_2$ 
and Ba$_{0.6}$K$_{0.4}$Fe$_2$As$_2$; and, more importantly for 
the investigations reported here, {\it (iii)} the report of local moments 
at room temperature ($T$) via Fe X-ray emission
spectroscopy~\cite{local,mannella}. 
Considering these experiments and others,
a better characterization of
the pnictides is that they are in the ``middle'' between the weak and strong 
Coulomb correlation limits~\cite{basov,rong,kotliar}. 
Because this intermediate Hubbard $U$ range is difficult for analytical approaches,
there is interest in the development of 
simpler models that can be studied via 
computational techniques to provide insight 
into such a difficult coupling regime.
The Hartree-Fock (HF) approximation to the Hubbard model~\cite{luo}
cannot be applied at room $T$ since HF approximations only lead 
to non-interacting fermions above the ordering temperature $T_{\rm N}$, and 
thus the local moment physics~\cite{local} cannot be reproduced~\cite{qmc}.

Recently, a Spin-Fermion (SF) model for the pnictides has been independently 
proposed by Lv {\it et al.}~\cite{Kruger} and Yin {\it et al.}~\cite{SF}. 
The model, very similar to those widely discussed for manganites, 
originally involved itinerant electrons in the $xz$ and $yz$ $d$-orbitals
coupled, via an on-site Hund interaction, to local spins (assumed classical) 
that represent the magnetic moment of 
the rest of the Fe orbitals (considered localized). 
The Hund interaction is supplemented by a nearest-neighbor (NN) and
next-NN (NNN) classical Heisenberg spin-spin interaction. 
This SF model has interesting features that makes it qualitatively suitable
for the pnictides,  particularly since by construction
the model has itinerant electrons in interaction with 
local moments~\cite{local,mannella} at all temperatures.

Phenomenological SF models have been proposed before 
for underdoped cuprates, with itinerant fermions representing
carriers locally coupled to classical spins representing 
the antiferromagnetic order parameter. These investigations unveiled 
stripe tendencies~\cite{SF-cuprates1},
ARPES and optical conductivity results~\cite{SF-cuprates2} 
similar to experiments, and even
the dominance of the $d$-wave channel in pairing~\cite{SF-cuprates3}. 
Thus, it is natural to apply now these ideas to the Fe superconductors.

As remarked already, SF models are also mathematically
similar to models used for the manganites~\cite{Dagotto}. 
Then, all the experience accumulated in the study 
of Mn-oxides can be transferred to the analysis of SF models 
for Fe-superconductors. In particular, one of our main objectives
is to study for the first time a SF model 
for pnictides employing Monte Carlo (MC)
techniques, allowing for an unbiased analysis of its 
properties. Moreover, to test the
model, challenging experimental results 
will be addressed. It is known that for
detwinned Ba(Fe$_{1-x}$Co$_x$)$_2$As$_2$
single crystals, a puzzling transport anisotropy has been discovered between the
ferromagnetic (FM) and antiferromagnetic (AFM) directions~\cite{Chu}.
In addition,
the resistivity vs. $T$ curves display
an unexpected peak at $T_{\rm N}$$\sim$130~K,
and the presumably weak effect of an applied uniaxial stress~\cite{Chu} still causes 
the anisotropy to persist well beyond $T_{\rm N}$. 
However, recent neutron results suggest that the transport anisotropy may be  
actually caused by strain effects that induce a shift
upwards of the tetragonal-orthorhombic 
and $T_{\rm N}$ transitions~\cite{wilson,blomberg}, 
as opposed to a spontaneous rotational symmetry-breaking state not induced
by magnetism or lattice effects. Then, theoretical guidance is needed.
While the low-$T$ anisotropy was already
explained as caused by the coupling between the spins and orbitals 
in the ${\bf Q}$=$(\pi,0)$ state~\cite{zhang}, 
the full transport curves at finite $T$ 
define a challenge that will be here addressed
for the nontrivial undoped case.

{\it Model and Method.} The SF model~\cite{Kruger,SF} is given by
\begin{equation}
H_{\rm SF} = H_{\rm Hopp} + H_{\rm Hund} + H_{\rm Heis}.
\end{equation}
The first term $H_{\rm Hopp}$ describes the Fe-Fe hopping of itinerant electrons. 
To better reproduce the band structure of pnictides~\cite{review},
three $d$-orbitals ($xz$, $yz$, $xy$) will be used instead of two.
The full expression for $H_{\rm Hopp}$
is cumbersome to reproduce here, but it is sufficient for the readers to
consult Eqs.(1-3) and Table I of Ref.~\onlinecite{three} for the mathematical form 
and the actual values of the hoppings in eV's. The density of 
relevance used here is $n$=4/3~\cite{three}. The Hund interaction 
is simply $H_{\rm Hund}$=$-J_{\rm H}\sum_{{\bf i}\alpha} 
{{{\bf S}_{\bf i}}\cdot{{{\bf s}^{\alpha}}_{\bf i}}}$, with ${\bf S}_{\bf i}$ the classical
spin at site ${\bf i}$ ($| {\bf S}_{\bf i}|$=1),
and ${{\bf s}^{\alpha}}_{\bf i}$ the itinerant-fermion spin of
orbital $\alpha$~\cite{comment-hund}. 
The last term $H_{\rm Heis}$ contains the spin-spin
interaction among the localized spins
$H_{\rm Heis}$=$J_{\rm NN} \sum_{\langle {\bf ij} \rangle} {{{\bf S}_{\bf i}}\cdot{{\bf S}_{\bf j}}}$
+ $J_{\rm NNN} \sum_{\langle\langle {\bf im} \rangle\rangle} {{{\bf S}_{\bf i}}\cdot{{\bf S}_{\bf m}}}$, where $\langle \rangle$ ($\langle \langle \rangle\rangle$) denotes NN (NNN) couplings. 
The particular ratio $J_{\rm NNN}$/$J_{\rm NN}$=2/3 was used in all the results below, leading
to $(\pi,0)$/$(0,\pi)$ magnetism~\cite{comment}. Any other ratio $J_{\rm NNN}$/$J_{\rm NN}$ larger
than 1/2 would have been equally suitable for our purposes. 

The well-known Monte Carlo (MC) technique for SF models~\cite{Dagotto}
will be here used to study $H_{\rm SF}$ at any $T$. In this technique, the
acceptance-rejection 
MC steps are carried out in the classical spins, while at each step
a full diagonalization of the fermionic sector (hopping plus on-site Hund terms)
for fixed classical spins is performed via library subroutines in order to
calculate the energy of that spin configuration. These frequent
diagonalizations render the technique CPU-time demanding.
The simulation is run on a finite 8$\times$8 cluster with 
periodic boundary conditions (PBC)
and uses the full $H_{\rm SF}$ model for the MC time evolution 
and generation of equilibrated configurations for the 
classical spins at a fixed $T$~\cite{MCsteps}. 
However, for the MC measurements those equilibrated configurations are assumed 
replicated in space but differing by a phase factor such that a better resolution
is reached with regards to the momentum ${\bf k}$. 
Since a larger lattice with more 
eigenstates gives a more continuous distribution of eigenenergies,
the procedure then reduces finite-size effects in the measurements.
This well-known method is often
referred to as ``Twisted'' Boundary Conditions (TBC)~\cite{salafranca}. 
In practice, phases $\Phi$ are added 
to the hopping amplitudes, schematically denoted as ``$t$'', at the boundary
via $t_{TBC}$=$e^{i \Phi} t$, with $\Phi$=$2\pi m/M$ ($m$=0,1,...,$M-1$) such that
the number of possible momenta in the $x$ or $y$ directions becomes $L$=8$\times$$M$.

\begin{figure}
\subfigure{\includegraphics[trim = 8mm 1mm 2mm 5mm,width=0.22\textwidth]{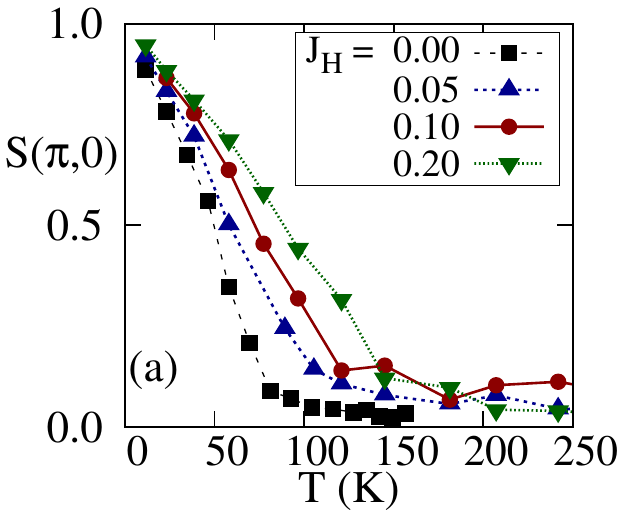}}
\subfigure{\includegraphics[trim = 5mm 1mm 5mm 5mm,width=0.22\textwidth]{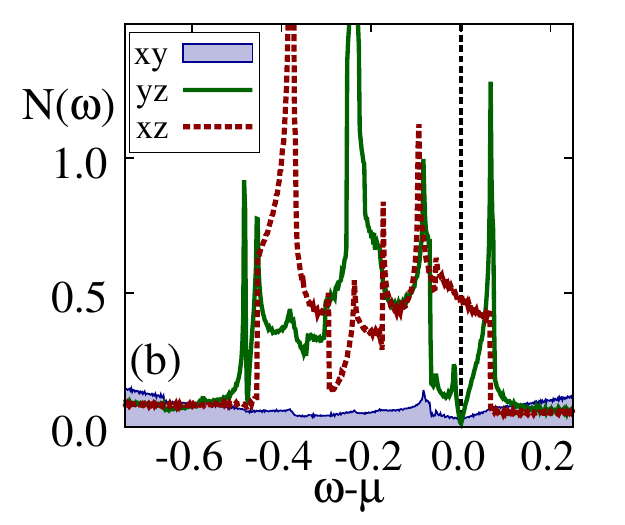}}
\subfigure{\includegraphics[trim = 8mm 2mm 2mm 5mm,width=0.22\textwidth]{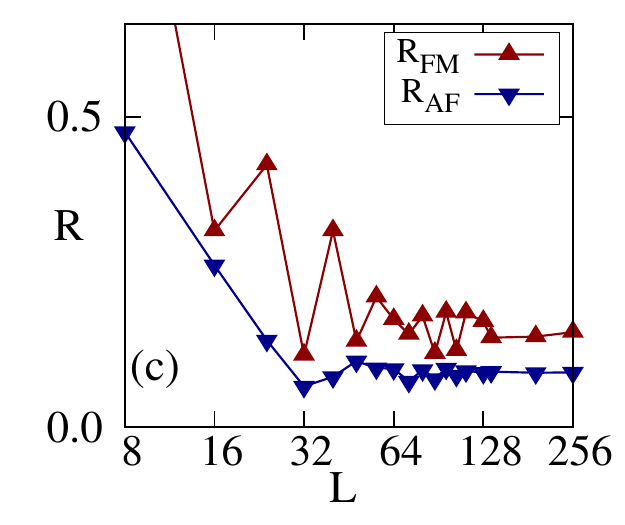}}
\subfigure{\includegraphics[trim = 5mm 2mm 5mm 5mm,width=0.22\textwidth]{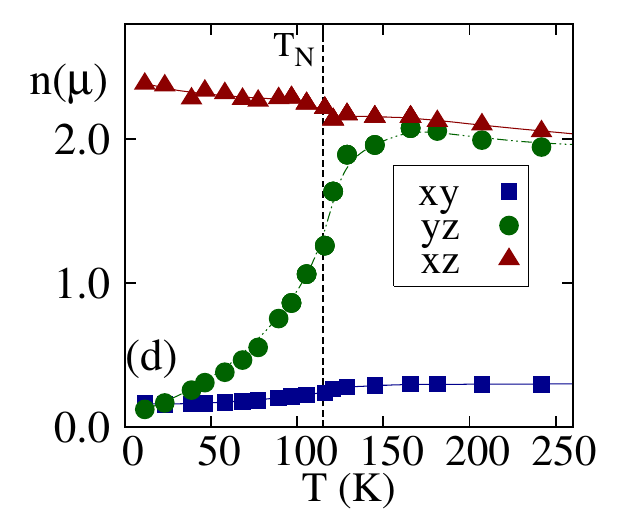}}
\caption{(color online) (a) Classical spins structure factor 
S($\pi$,0) (normalized to 1) vs. $T$, for the $J_{\rm H}$'s 
indicated, using the PBC 8$\times$8 cluster and  $J_{\rm NN}$=0.015. 
The oscillations in the data are indicative of the error bars. (b) Density of
states $N(\omega)$ of each orbital ($\mu$= chemical potential), 
using TBC with $L$=512, at $T$=0 K and $J_{\rm H}$=0.1~eV, for the
perfect $(\pi,0)$ magnetic state.
(c) Resistance $R$ vs. $L$ (TBC 8$\times$8) for the FM and AFM directions
of the perfect $(\pi,0)$ magnetic state ($J_{\rm H}$=0.1~eV). 
 (d)  The occupation at the FS $n(\mu)$ (see text) of the three orbitals
vs.~$T$, using $L$=256. A coupling $J_{\rm NN}$=0.016 (0.014) along the $x$ ($y$) 
axis was used (see text).
\label{fig1}}
\vskip -0.5cm
\end{figure}

{\it Results.} 
Figure~\ref{fig1}(a) contains MC results for the structure 
factor of the classical spins, defined 
as $S({\bf q})$=${1}\over{N^2}$$\sum_{{\bf i},{\bf j}} 
\langle {{\bf S}_{\bf i}}\cdot{{\bf S}_{\bf j}} 
\rangle e^{i{\bf q}\cdot{({\bf i-j}})}$ ($N$  = number of sites), illustrating
the development of ${\bf Q}$=$(\pi,0)$ magnetic order as $T$ is reduced. 
Since a ratio $J_{\rm NNN}$/$J_{\rm NN}$$>$1/2
is ``frustrating'', finding ${\bf Q}$-order at $J_{\rm H}$=0 is not surprising, but  
Fig.~\ref{fig1} shows that this order remains stable turning on $J_{\rm H}$ in the
range investigated, as opposed to inducing transitions to other states. 
The chosen value of $J_{\rm NN}$ in Fig.~\ref{fig1}(a) 
leads to a $T_{\rm N}$ similar to that in BaFe$_2$As$_2$. 
The low-$T$ orbitally-resolved 
electronic density-of-states (DOS) is in Fig.~\ref{fig1}(b). The
${\bf Q}$ magnetic order opens a pseudogap (PG) in the $yz$ orbital, while
the others are not much affected. This PG generation was previously discussed 
when contrasting theory and ARPES experiments~\cite{weight} and it  should not be confused
with long-range orbital-order, that in this SF model occurs
at $J_{\rm H}$$\sim$0.4 or larger.

Figure \ref{fig1}(c) contains the evolution of the 8$\times$8-cluster
resistance $R$ increasing the number of momenta via the TBC, calculated
via standard procedures~\cite{Dagotto,Verges}. While the
ratio of $R$'s along the FM and AFM directions 
is always $>1$, i.e. qualitatively correct, 
TBC with $L$=256 is needed to reach
stable values.
In addition, the occupation 
of the three orbitals at the FS (Fig.~\ref{fig1}(d)) 
was defined as $n(\mu)$=$\int d\epsilon n(\epsilon) 
\beta e^{\beta (\epsilon - \mu)}/(1+e^{\beta(\epsilon - \mu)})^2$, involving
the $\mu$-derivative of the fermionic population.
As $T$ increases and the ${\bf Q}$ order weakens, 
the $xz$-$yz$ orbitals populations converge to the same values.

The results of Fig.~\ref{fig1}, and others below,
were obtained introducing a ``small'' explicit
asymmetry along the $x$ and $y$ axes for $J_{\rm NN}$, namely
a generalized $J^{\alpha}_{\rm NN}$ ($\alpha$=$x,y$) was used. Its purpose
is to mimic the orthorhombic distortion and strain effects~\cite{cruz,wilson,blomberg} 
and judge if the present calculations reproduce 
transport experiments~\cite{Chu,wilson,blomberg}. Consider the
ratio $r_{\rm NN}$=$J^x_{\rm NN}$/$J^y_{\rm NN}$. 
Using the dependence
of the hopping amplitudes with the distance $u$ between $d$- and $p$-orbitals 
i.e. $t_{pd}$$\sim$$1/u^{7/2}$~\cite{harrison}, the angles involved in the Fe-As-Fe bonds,
the low-$T$ lattice parameters~\cite{cruz}, 
fourth-order perturbation in the hoppings for 
the Fe-Fe superexchange, and, more importantly, 
neglecting contributions of the $yz$ orbital
that is suppressed at the Fermi level~\cite{weight} as long as $(\pi,0)$
spin fluctuations dominate leads to a 
crude estimation $r_{\rm NN}$$\sim$1.4. Since this is likely an upper bound,
the ratio used in our MC simulation 
$r_{\rm NN}$$\sim$1.14, assumed to be temperature independent, is reasonable. 
Other crude estimations
including the direct Fe-Fe hoppings~\cite{harrison} or employing the
lattice parameters under pressure~\cite{wilson} lead to similar ratios. 
Then, our asymmetry value is qualitatively realistic.

\begin{figure}
\includegraphics[width=0.42\textwidth]{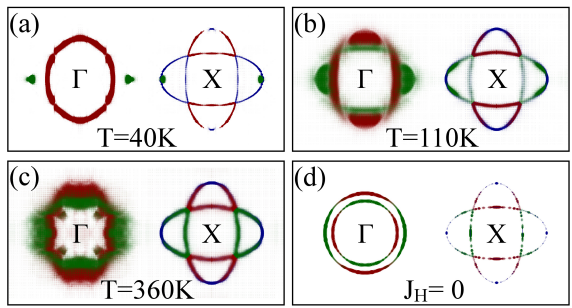}
\vskip -0.1cm
\caption{(color online) 
$A({\bf k},\omega)$ at $\omega$=$\mu$ (TBC 8$\times$8 $L$=512). The
model used (see Ref.~\onlinecite{three}) includes the staggered As modulation
out of the FeAs layer. Thus, our results are in the {\it folded} Brillouin zone convention,
and for this reason two electron pockets (as opposed to just one) are centered at X in
the panels above. 
The pocket elongated vertically at X would corresponds to 
a pocket at Y in the unfolded convention if the As modulation is considered
via a quasicrystal momentum~\cite{three}. 
(a-c) are for  $J_{\rm H}$=0.1~eV. Red, green, and blue are for the
$xz$, $yz$, and $xy$ orbitals, respectively. 
(a) $T$=$40K$, below $T_{\rm N}$. The $(\pi,0)$ magnetic order 
induces $yz$-orbital electron satellite pockets.
(b) $T$=$110K$$\sim$$T_{\rm N}$.
In this regime, the MC configurations display small 
coexisting patches of $(\pi,0)$ and $(0,\pi)$ order (see text), 
creating almost symmetric $xz$ and $yz$ features around $\Gamma$.
(c) Large $T$=$360$~K, with no remnants of the $(\pi,0)$ order.
The FS becomes a broaden version of the non-interacting FS at $J_{\rm H}$=$0$,
shown in (d) at $T$=100~K.
\label{fig2}}
\vskip -0.5cm
\end{figure}

Previous investigations~\cite{luo} showed that  
the $T$=0 HF approximation to the undoped multiorbital Hubbard model 
can reproduce neutron diffraction results and ARPES data. A similar degree
of accuracy should be expected from any reasonable model for the pnictides,
including $H_{\rm SF}$. To test this assumption, 
the one-particle spectral function $A({\bf k},\omega)$ 
was calculated, and the FS at different $T$'s is shown in Fig.~\ref{fig2},
contrasted against the low-$T$ fermionic non-interacting limit $J_{\rm H}$=$0$.
At low-$T$ in the ordered state  the expected 
asymmetry between the $(\pi,0)$ and $(0,\pi)$ electron 
pockets is observed (not shown), and more importantly 
satellite pockets (with electron character) develop close to the $\Gamma$ hole
pockets (Fig.~\ref{fig2}(a)), as in ARPES experiments~\cite{luo,arpes}. 
Thus, the SF model studied here passes the low-$T$ ARPES test.
As $T$ increases, at or well above
$T_{\rm N}$ (Figs.~\ref{fig2}(b) and (c)) 
the $xz$ and $yz$ differences are reduced and rotational
invariance is recovered, albeit
with a FS broader than in the non-interacting
low-$T$ limit (Fig.~\ref{fig2}(d)).

\begin{figure}
\includegraphics[width=0.4\textwidth]{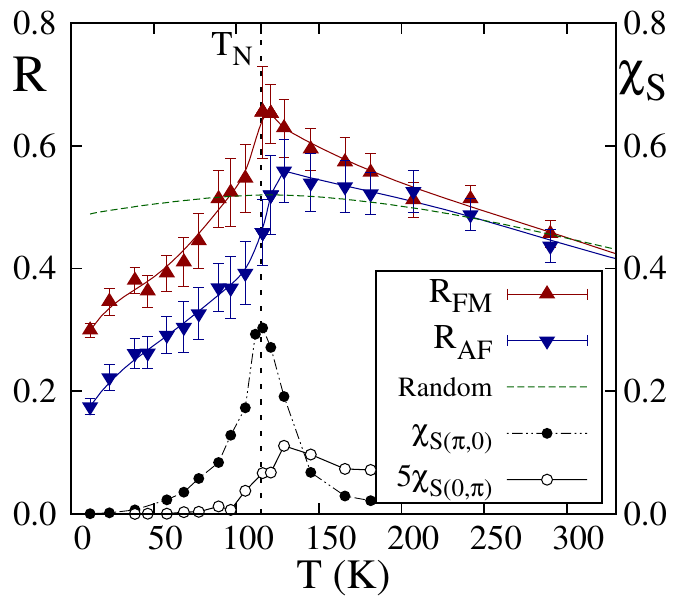}
\vskip -0.2cm
\caption{(color online) 
Resistance $R$ of the SF model calculated via MC simulations, at $J_{\rm H}$=0.1~eV
and using the $L$=256 TBC 8$\times$8 cluster. $T_{\rm N}$ 
is indicated, and the magnetic susceptibilities $\chi_S$ are also shown. The
FM-AFM directions asymmetry 
is evident at low $T$ (note that the FM and AFM labels 
simply refer to the $y$ and $x$ directions, respectively,
and not to fully FM or AFM spin configurations). 
As $T$ increases the symmetry is restored, and
the curves display a peak at $T_{\rm N}$. 
A small symmetry-breaking
difference $r_{\rm NN}$=1.14 is used (see text).
In green (dashed line) are the results
for random spin configurations, showing that their $R$ is
smaller than in the MC results near $T_{\rm N}$. The width of the $\chi$ peaks
extend to $\sim$1.5$T_{\rm N}$, in agreement with neutron scattering 
for CaFe$_2$As$_2$~\cite{diallo}.
\label{fig3}}
\vskip -0.5cm
\end{figure}

{\it $R$ vs. $T$ curves.} 
Our most important result is the $T$ dependence of $R$ along the two axes,
shown in Fig.~\ref{fig3}. It is visually clear that these results
are similar to the transport data of Ref.~\onlinecite{Chu}, 
particularly
after realizing that lattice effects, that cause the continuous raise of 
$R$ with $T$ in the experiments, 
are not incorporated in the SF model.  
A clear difference exists between the FM and AFM 
directions at low $T$, induced by the
$(\pi,0)$ magnetic order that breaks spontaneously 
rotational invariance.
At low $T$, this difference was understood 
in the Hubbard-model HF approximation~\cite{zhang}
based on the reduction of the $yz$ orbital population (Fig.~\ref{fig1}(d)).
This explanation
is equally valid in the SF model,
and at low $T$ the SF model and the Hubbard model, 
when treated via the HF approximation, lead to similar physics. 

The most interesting result in Fig.~\ref{fig3} is the development
of a peak at $T_{\rm N}$, and the subsequent slow convergence of $R$ toward 
similar values along both
directions with further increasing $T$ (as already discussed, to model  better
the effect of uniaxial stress~\cite{Chu}, a weak symmetry-breaking difference
between the NN Heisenberg couplings along $x$ 
and $y$ was included).
To our knowledge, this is 
the first time that the full $R$ vs. $T$ curve is successfully 
reproduced via computational studies.

\begin{figure}
\includegraphics[width=0.4\textwidth]{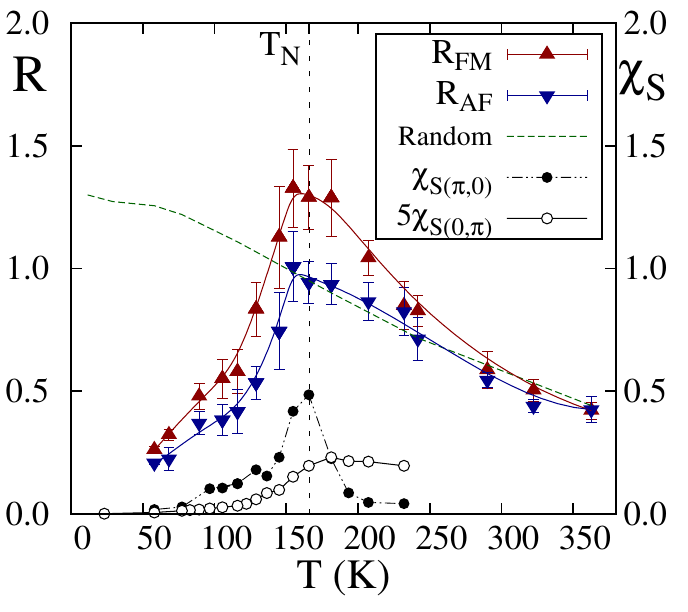}
\vskip -0.3cm
\caption{(color online) 
Resistance $R$ calculated using a $L$=64 TBC 16$\times$16 cluster, 
with $J_{\rm NN}$=0.032 (0.028) along the $x$ ($y$) axis. 
The spin configurations are generated via $H_{\rm Heis}$, while
the $R$ measurements use the full $H_{\rm SF}$, 
at $J_{\rm H}$=0.1~eV. The rest of the notation is as in Fig.~\ref{fig3}.
\label{fig4}}
\vskip -0.5cm
\end{figure}

A study as in Fig.~\ref{fig3} using a larger cluster, 
e.g. 16$\times$16, is not practical since the
computer time grows like $N^4$ ($N$=number of sites), leading to
an effort 256 times larger.
However, results as in Fig.~\ref{fig1}(a) 
indicate that the classical spins configurations generated
merely by the spin-spin interaction could
be qualitatively similar to those generated by
the full SF model, as long as $J_{\rm H}$ does not push the system into
a new phase. Thus, the MC evolution could
be carried out with $H_{\rm Heis}$ only, while measurements 
can still be performed using the full diagonalization
of $H_{\rm SF}$. Such measurements (very CPU-time consuming) must be
sufficiently sparsed in the MC evolution to
render the process practical.
This procedure was implemented 
on a TBC 16$\times$16 cluster, with $L$=64~\cite{comment2}.
The results for $R$ are in Fig.~\ref{fig4}, and they show a remarkable
similarity with Fig.~\ref{fig3}, and with experiments. 
Thus, the essence of the $R$ vs. $T$ 
curves is  captured by electrons moving in  the spin configurations generated by 
$H_{\rm Heis}$. Size effects are small in the range analyzed here.

\begin{figure}
\subfigure{\includegraphics[trim = 20mm 1mm 25mm 5mm,width=0.22\textwidth]{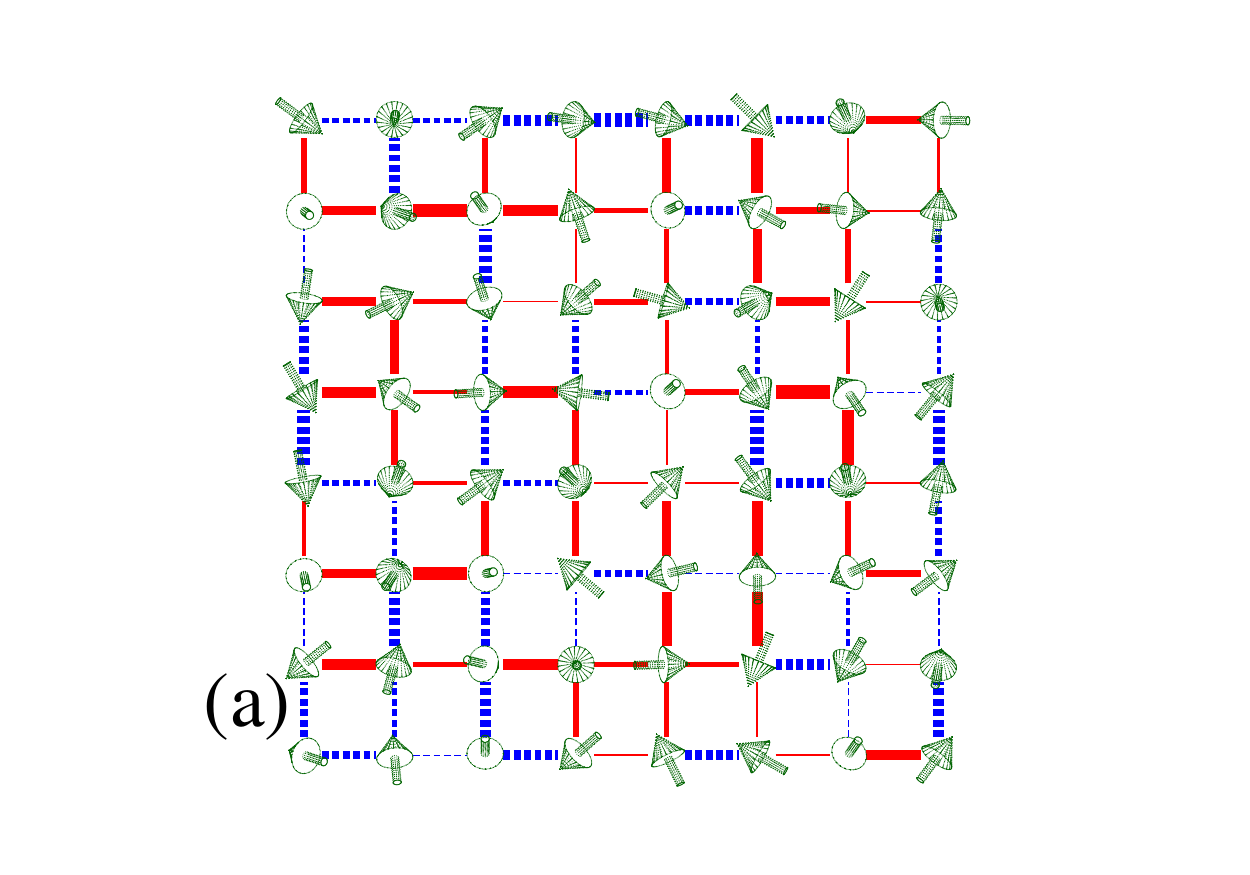}}
\subfigure{\includegraphics[trim = 25mm 1mm 20mm 5mm,width=0.22\textwidth]{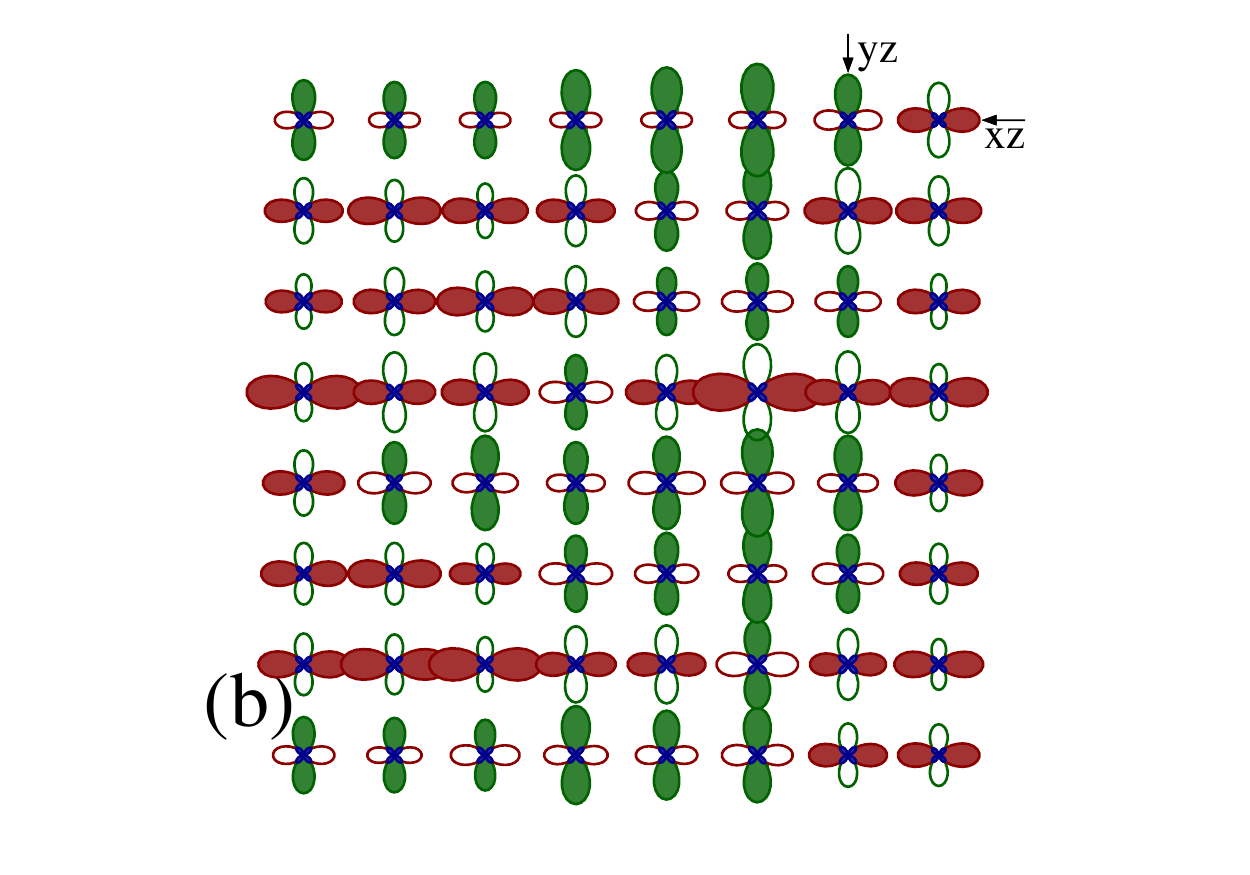}}
\vskip -0.25cm
\caption{(color online) (a) Typical MC-generated classical spin configuration 
on an 8$\times$8 cluster at $T_{\rm N}$$\sim$110~K, $J_{\rm H}$=0.1~eV, 
and $J_{\rm NN}$=0.016 (0.014)
along the $x$ ($y$) axis. 
The red (blue) lines denote AFM (FM) NN correlations, of intensity proportional 
to the width. (b) The dominant orbital at the FS at each site for the configuration
used in (a), calculated using $n(\mu)$ as in Fig.~1(d). Red (green) denote the $xz$ ($yz$) orbital, 
with a size proportional to the density.
\label{fig5}}
\vskip -0.5cm
\end{figure}

What causes the increase of $R$ upon cooling before $T_{\rm N}$ is reached, 
displaying insulating characteristics? Since 
our results are similar to experiments, an analysis
of the MC-equilibrated configurations may provide qualitative insight into 
their origin. In Fig.~\ref{fig5}(a),
a typical MC
configuration of classical spins is shown. 
The colors at the links illustrate the relative orientation of the two spins 
at the ends. The $(\pi,0)$ long-range order is lost, but individual
spins are not randomly oriented. In fact, the state contains 
small regions
resembling locally either a $(\pi,0)$ or $(0,\pi)$ order 
(short-range spin order), and
$S({\bf q})$ still displays broad
peaks at those two wavevectors.
In standard mean-field approximations 
there are no precursors of the magnetic 
order above $T_{\rm N}$, but 
in the SF model there are short-range fluctuations in the
same regime. 

States with ``spin patches'' as in Fig.~\ref{fig5}(a)
lead to a concomitant patchy orbital order at the FS shown in
Fig.~\ref{fig5}(b), where the most populated 
orbital at $\mu$ at every site, either $xz$ or $yz$, 
are indicated. 
The orbital orientation suggests
that the $xz$ ($yz$) 
FS population favors transport along the $x$ ($y$) axis. The patchy order
should have a resistance {\it larger} 
than that of a randomly-oriented spin background. This is confirmed 
by calculating $R$ vs. $T$ in the absence
of a guiding Hamiltonian, i.e. by generating random spin configurations. 
The results are in Figs.~\ref{fig3} and
\ref{fig4} (green dashes) and 
their values are indeed 
{\it below} those of the peak resistance at $T_{\rm N}$ 
of the full SF model, i.e. with configurations as in  Fig.~\ref{fig5}. Then,
the effect of strain coupled to short-range spin and FS orbital order appears 
to be 
the cause of the peak in the $R$ vs. $T$ curves~\cite{fernandez,comment-asym}. 
Using a smaller (but nonzero) anisotropy, 
the $R$-$T$ curves display a concomitantly smaller anisotropy, but 
still they have a small peak at $T_{\rm N}$ (not shown). 
Thus, the patchy states may also 
explain the insulating properties of Fe$_{1.05}$Te~\cite{FT} 
and (Tl,K)Fe$_{2-x}$Se$_2$~\cite{KFS} above $T_{\rm N}$.

{\it Conclusions.} The Spin-Fermion model 
for pnictides was here studied with MC techniques. 
The magnetic and ARPES properties of the undoped compounds
are well reproduced. Including a small explicit 
symmetry breaking to account for strain effects,
the resistance $R$ vs. $T$ curves are qualitatively
similar to those observed for BaFe$_2$As$_2$~\cite{Chu}, 
including a peak at $T_{\rm N}$ that appears
caused by short-range spin and FS orbital order. In our calculations, 
the anisotropy above $T_{\rm N}$ exists {\it only} as long as a strain distortion exists,
compatible with results for annealed BaFe$_2$As$_2$ samples~\cite{pnas}. 
This successful application of a SF model paves 
the way to more demanding efforts involving doped systems
where anisotropy effects are stronger 
than in the undoped limit~\cite{Chu}.

{\it Acknowledgment.} The authors thank I.R. Fisher,
M. Daghofer, S. Dong, W. Lv, S. Wilson, 
X. Zhang, Q. Luo, and A. Nicholson for useful
discussions. This work was supported by the US Department of Energy,
Office of Basic Energy Sciences, Materials Sciences
and Engineering Division, and also by the National Science
Foundation under Grant No. DMR-11-04386 (S.L., C.S., A.M., E.D.). 
G.A. was supported by the Center for
Nanophase Materials Sciences, sponsored by the Scientific User Facilities 
Division, BES, U.S. DOE. This research used resources of
the National Center for Computational Sciences, as well as the OIC at ORNL.

\bibliographystyle{prsty-etal}

\begin{thebibliography}{1}


\bibitem{review} D. C. Johnston, Adv. Phys. {\bf 59}, 803 (2010).

\bibitem{cruz} C. de la Cruz, Q. Huang, J. Lynn, I. Ratcliff, 
J. Zaretsky, H. Mook, G. Chen, J. Luo, N. Wang, and P. Dai, 
Nature (London) {\bf 453}, 899 (2008).

\bibitem{spinIC} D. K. Pratt, 
M. G. Kim, A. Kreyssig, Y. B. Lee, G. S. Tucker, A. Thaler, W. Tian, 
J. L. Zarestky, S. L. Bud'ko, P. C. Canfield, B. N. Harmon, A. I. Goldman, 
and R. J. McQueeney, 
Phys. Rev. Lett. {\bf 106}, 257001 (2011).

\bibitem{davis} T.-M. Chuang, 
M. P. Allan, J. Lee, Y. Xie, N. Ni, S. L. Bud'ko, G. S. Boebinger, P. C. Canfield, and J. C. Davis, 
Science {\bf 327}, 181 (2010).

\bibitem{shimojima} T. Shimojima, F. Sakaguchi, K. Ishizaka, Y. Ishida, T. Kiss, 
M. Okawa, T. Togashi, C.-T. Chen, S. Watanabe, 
M. Arita, K. Shimada, H. Namatame, M. Taniguchi, K. Ohgushi, S. Kasahara, T. Terashima, 
T. Shibauchi, Y. Matsuda, A. Chainani, and S. Shin, 
Science {\bf 332}, 564 (2011).

\bibitem{local} H. Gretarsson, A. Lupascu, J. Kim, D. Casa, T. Gog, 
W. Wu, S. R. Julian, Z. J. Xu, J. S. Wen, G. D. Gu, 
R. H. Yuan, Z. G. Chen, N.-L. Wang, S. Khim, K. H. Kim, M. Ishikado, I. Jarrige, S. Shamoto, 
J.-H. Chu, I. R. Fisher, and Y-J. Kim, 
Phys. Rev. B {\bf 84}, 100509(R) (2011).

\bibitem{mannella} F. Bondino, E. Magnano, M. Malvestuto, F. Parmigiani, 
M. A. McGuire, A. S. Sefat, B. C. Sales, R. Jin, D. Mandrus, E. W. Plummer, 
D. J. Singh, and N. Mannella, 
Phys. Rev. Lett. {\bf 101}, 267001 (2008).

\bibitem{basov} M. M. Qazilbash, J. J. Hamlin, R. E. Baumbach, Lijun Zhang, 
D. J. Singh, M. B. Maple, and D. N. Basov, 
Nat. Phys. {\bf 5}, 647 (2009).

\bibitem{rong} R. Yu, K.T. Trinh, A. Moreo, M. Daghofer, J. Riera, S. Haas, 
and E. Dagotto, 
Phys. Rev. B {\bf 79}, 104510 (2009). 

\bibitem{kotliar} Z.P. Yin, K. Haule, and G. Kotliar, Nat. Mat. {\bf 10}, 932 (2011).

\bibitem{luo} 
Q. Luo, G. Martins, D.-X. Yao, M. Daghofer, R. Yu, A. Moreo, and E. Dagotto, 
Phys. Rev. B {\bf 82}, 104508 (2010). 

\bibitem{qmc} 
Determinantal MC 
cannot be applied to the multiorbital case due
to the ``sign problem'' even in the undoped limit.

\bibitem{Kruger} W. Lv, F. Kr\"uger, and P. Phillips, 
Phys. Rev. B {\bf 82}, 045125 (2010). 

\bibitem{SF} 
W.-G. Yin, C.-C. Lee, and W. Ku, 
Phys. Rev. Lett. {\bf 105}, 107004 (2010).




\bibitem{SF-cuprates1} 
C. Buhler, S. Yunoki, and A. Moreo, Phys. Rev. Lett. {\bf 84}, 2690, (2000).


\bibitem{SF-cuprates2} M. Moraghebi, C. Buhler, S. Yunoki and A. Moreo, 
Phys. Rev. B {\bf 63}, 214513 (2001); M. Moraghebi, S. Yunoki and A. Moreo,  
Phys. Rev. B {\bf 66}, 214522 (2002). 
 
\bibitem{SF-cuprates3} 
M. Moraghebi, S. Yunoki, and A. Moreo, Phys. Rev. Lett. {\bf 88}, 187001 (2002). 

\bibitem{Dagotto} E. Dagotto, T. Hotta, and A. Moreo, 
Phys. Rep. {\bf 344}, 1 (2001). 

\bibitem{Chu}
J-H. Chu, J. G. Analytis, K. De Greve, P.L. McMahon, 
Z. Islam, Y. Yamamoto, and I.R. Fisher
Science {\bf 13},  824 (2010). See also I.R. Fisher, L. Degiorgi, and 
Z.X. Shen, Rep. Prog. Phys. {\bf 74}, 124506 (2011).

\bibitem{wilson}
C. Dhital, Z. Yamani, Wei Tian, J. Zeretsky, A. S. Sefat, 
Ziqiang Wang, R. J. Birgeneau, and S. D. Wilson,
Phys. Rev. Lett. {\bf 108}, 087001 (2012).

\bibitem{blomberg} E. C. Blomberg, A. Kreyssig, M. A. Tanatar, R. Fernandes, M. G. Kim, 
A. Thaler, J. Schmalian, S. L. Bud'ko, P. C. Canfield, A. I. Goldman, and R. Prozorov,
Phys. Rev. B {\bf 85}, 144509 (2012).

\bibitem{zhang} X. Zhang and E. Dagotto, Phys. Rev. B {\bf 84}, 132505 (2011).

\bibitem{three} M. Daghofer, A. Nicholson, A. Moreo, and E. Dagotto, 
Phys. Rev. B {\bf 81}, 014511 (2010). 

\bibitem{comment-hund} Although the fermions do not have a direct Hund interaction
among themselves, an effective one is generated via the interaction with the classical variables
(more formally, ``integrating out'' the classical spins should induce a Hund coupling among the fermions).


\bibitem{comment} The transition
from $(\pi,\pi)$ order to $(\pi,0)$/$(0,\pi)$ order occurs at 
$J_{\rm NNN}$/$J_{\rm NN}$=1/2, within $H_{\rm Heis}$.

\bibitem{MCsteps} Typically 20,000 MC sweeps through the lattice
are used for thermalization followed by 10,000 for the measurements that occur every
20 configurations.



\bibitem{salafranca} J. Salafranca, G. Alvarez, and E. Dagotto, 
Phys. Rev. B {\bf 80}, 155133 (2009).

\bibitem{weight} M. Daghofer, Q.-L. Luo, R. Yu, D. X. Yao, A. Moreo and E. Dagotto, 
Phys. Rev. B {\bf 81}, 180514(R) (2010).

\bibitem{Verges}
J. A. Verg\'es, 
Comput. Phys. Commun. {\bf 118}, 71 (1999).

\bibitem{harrison} W.A. Harrison, {\it Electronic Structure and the Properties
of Solids}, (Dover, 1989).

\bibitem{arpes} M. Yi, D.H. Lu, J.G. Analytis, J.-H. Chu, S.-K. Mo, R.-H. He,
M. Hashimoto, R. G. Moore, I.I. Mazin, D.J. Singh, Z. Hussain,
I.R. Fisher, and Z.-X. Shen, 
Phys. Rev. B {\bf 80}, 174510 (2009).

\bibitem{diallo} S.O. Diallo, D.K. Pratt, R.M. Fernandes, W. Tian, J.L. Zarestky, M. Lumsden, 
T.G. Perring, C.L. Broholm, N. Ni, S.L. Bud'ko, P.C. Canfield, 
H.-F. Li, D. Vaknin, A. Kreyssig, A.I. Goldman, and R.J. McQueeney,
Phys. Rev. B {\bf 81}, 214407 (2010).


\bibitem{comment2} $\sim$200,000 ($\sim$100,000) MC steps for thermalization (measurements) 
were employed. The $R$ calculations are very CPU time consuming, 
thus only $\sim$20 were performed, but self-averaging effects render the error bars small.

\bibitem{fernandez} The resistivity anisotropy has also been recently explained
in similar terms via a nematic phase above $T_{\rm N}$ 
(R.M. Fernandes, E. Abrahams, and J. Schmalian,
Phys. Rev. Lett. {\bf 107}, 217002 (2011)). The existence of a nematic phase
in the SF model will be studied in the future.

\bibitem{comment-asym} This
qualitative argumentation should be substantiated by 
transport calculations involving, e.g., scattering rates.

\bibitem{FT} G. F. Chen, Z. G. Chen, J. Dong, W. Z. Hu, G. Li, 
X. D. Zhang, P. Zheng, J. L. Luo, and N. L. Wang,
Phys. Rev. B {\bf 79}, 140509(R) (2009).

\bibitem{KFS}M. H. Fang, H. D.Wang, C. H. Dong, Z. J. Li, C. M. Feng, 
J. Chen, and H. Q. Yuan,
EPL {\bf 94}, 27009 (2011).

\bibitem{pnas} M. Nakajima, T. Liang, S. Ishida, Y. Tomioka, K. Kihou, 
C. H. Lee, A. Iyo, H. Eisaki, T. Kakeshita, T. Ito, and S. Uchida,
PNAS {\bf 108}, 12238 (2011).


\end{thebibliography}

\end{document}